\documentclass[aip,amsmath,amssymb,graphicx,preprint]{revtex4-1}

\usepackage{graphicx}
\usepackage{dcolumn}
\usepackage{bm}

\usepackage[utf8]{inputenc}
\usepackage[T1]{fontenc}
\usepackage{mathptmx}
\usepackage{etoolbox}

\draft 

\makeatletter
\def\@email#1#2{%
 \endgroup
 \patchcmd{\titleblock@produce}
  {\frontmatter@RRAPformat}
  {\frontmatter@RRAPformat{\produce@RRAP{*#1\href{mailto:#2}{#2}}}\frontmatter@RRAPformat}
  {}{}
}%
\makeatother

\begin{document}

\preprint{}

\title{All field emission experiments are noisy, ... are any meaningful ?} 



\author{Anthony Ayari}
\author{Pascal Vincent}
\author{Sorin Perisanu}
\author{Philippe Poncharal} 
\author{Stephen T. Purcell}
\email[]{anthony.ayari@univ-lyon1.fr.}
\affiliation{Univ Lyon, Univ Claude Bernard Lyon 1, CNRS, Institut Lumi\`ere Mati\`ere, F-69622, VILLEURBANNE, France.}


\date{\today}

\begin{abstract}
Representing field emission data on a Fowler-Nordheim plot is both very common and strongly not recommended. It leads to a spurious estimation of the emitter parameters despite a very good data fit. There is a lack of a reliable method of analysis and a proper estimation of the uncertainty in the extracted parameters. In this article, we show that the uncertainty in the estimation of the field enhancement factor or the emission area can be as high as $\pm 50\%$ even for a tungsten single emitter in good ultra-high vacuum conditions analyzed by the Murphy-Good model. Moreover, the choice of the exact Murphy-Good method can have a noticeable impact. We found that advanced analysis methods, based on the measurement of the differential conductance of the emitter, are so demanding in terms of emitter stability that up to now its requirements are probably out of reach in any field emission laboratory.
\end{abstract}


\maketitle 

\section{Introduction}

Vacuum electron sources have been at the heart of important applications such as electron microscopy, radio communication and electronics during the 20$^{th}$ century. This led to extensive studies of the main electron emission mechanisms : thermionic emission,photo-emission, secondary emission and field emission\cite{jensen2017introduction}. By its very nature, field emission is probably the most intriguing process of the four. It is one of the first physical mechanisms involving the tunneling effect and is described by the so-called Fowler-Nordheim (FN) theory\cite{fowler1928electron}. This model has had a great influence in the vacuum source community but also in fields as varied as molecular electronics\cite{beebe2006transition}, chemistry\cite{bell2013tunnel} or biology\cite{devault1967electron}. Despite its notoriety, the FN model is no longer recommended by the field emission community\cite{forbes202121st}. Furthermore, this model and its more exact version based on the Murphy-Good\cite{murphy1956thermionic} theory (MG) has never been fully experimentally verified. We give in this article a brief overview of the past and recent attempts to close the gap between field emission theories and experiments. Then a set of experimental I-V curves from a tungsten emitter are extensively analyzed and the uncertainty in the extraction of physical parameters is estimated. The origin of these uncertainties is identified by determining the experimental noise, the current drift and comparison with various analysis methods. Finally, numerical simulations are performed in order to highlight the experimental difficulties and challenges to get more reliable field emission studies.


\section{A brief history of experimental verification of field emission theory}

The misuse of field emission theory is probably as old as the questioning about its validity. There exists abundant literature on the subject and sometimes criticisms about the work of other researchers coexist in the same article with misuse of the theory by the author itself. In order to escape this pitfall, we will focus on what should not be done and cowardly avoid proposing a good method to test field emission.

\subsection{The nightmare of the three parameters for field emission}
For many experimentalists, it is well known that field emission experiments can be interpreted thanks to the FN theory\cite{fowler1928electron} which predicts that the current I is given by :

\begin{equation}\label{eqFN}
    I = aS\frac{(\beta V)^2}{\phi}\exp(-b\frac{\phi^{3/2}}{\beta V})
\end{equation}

where a and b are terms that depend only on universal physical constants, S is the emission area, $\beta$ is the field enhancement factor relating the applied voltage V to the electric field and $\phi$ is the work function. Then plotting the current in Fowler-Nordheim coordinates, (i.e. representing the logarithm of $I/V^2$ as a function of $1/V$) gives a straight line over several orders of magnitude. The goal is then to hope that an extraordinary value will be extracted from the slope and intercept of a linear fit, like a huge current density or $\beta$. Probably the most impressive attempt and the best example of the lack of reliability of this method can be found in ref. \onlinecite{feist1968cold} p.18. where a work function as ridiculously low as 0.01 eV was obtained. This is a point that is still too much ignored: there is no point in trying to publish about exceptionally low work function as this world record will be too hard to beat. One of the problems comes from the fact that Eq. \ref{eqFN} needs three parameters but only two (the slope and the intercept) can be extracted from the FN plot. It is then necessary to guess one of the parameters and often this guess is incorrect. Another issue in Eq. \ref{eqFN} is that it was obtained for a triangular tunneling barrier whereas the image charge has a huge effect on the barrier transparency. Using Eq. \ref{eqFN} will lead to an overestimation of the area by several orders of magnitude. It is then rather well known that MG theory\cite{murphy1956thermionic} is more appropriate. In this article, for simplicity, other improvements of the model will be disregarded such as the replacement of the smooth flat surface hypothesis by a proper atomistic structure, the field and temperature dependence of the work function or the calculation of the electron density beyond the Sommerfeld free electron model.

Even with a more appropriate theory, many problems remain. For instance, it is sometimes considered that it is safer to take the work function as an input parameter and then to deduce S and $\beta$. The main reason is that a reasonable range of work function is between 3.5 and 6 eV whereas $\beta$ and S are related to the radius of curvature of the emitter and this radius can commonly vary between roughly one nanometer to several hundreds of nanometers. Nevertheless, some uncertainty in the work function remains and this uncertainty can have an important impact on the estimation of the other parameters. An interesting example was shown by Muller\cite{muller1955work} where even in the apparently simple case of a 110 facet of an ultraclean field emitter the work function might differ by 0.5 eV depending on the annealing conditions (i.e. 1200$^\circ$C versus 2200$^\circ$C). To make this worse, it is highly unlikely that S, $\beta$ and $\phi$ are constant over the entire voltage range (see for instance ref.\onlinecite{jensen2017introduction} chap. 30).

\subsection{The longest IV curve}

The most certain fact about field emission is that the current is dominated by an exponential term and a convincing proof has been given by  Dyke and Trolan\cite{dyke1953field} over seven orders of magnitude in current. However, It is more difficult to follow them when they claim that "The wave mechanical, image force corrected theory quantitatively predicted the observed average current density up to that density for which space charge dominated the emission." Although they provide numerical values for the field enhancement factor or the current density, it is hard to estimate if these values are quantitatively correct even when some margin of error of the order of 15 $\%$ is given. Their linear curves are not plotted in FN coordinates, they take into account the change of emission area with the voltage and despite their effort in terms of vacuum conditions (estimated below 10$^{-12}$ Torr), it can be seen in log scale that their current data points are not perfectly reproducible, with a difference in current sometimes higher than 10$\%$ and a deviation with the theoretical current density higher than 25 $\%$.

\subsection{From qualitative to quantitative estimations}

As it seems impossible to obtain reliable information from an I-V curve only, additional methods have been proposed. It was for instance proposed to perform a joint measurement of the current and the electron energy distribution\cite{young1962progress,young1966effect}. The width of the peak or the slope of the side below the Fermi energy should theoretically give an independent estimate of a relationship between $\phi$ and $\beta$. Although there isn't a clear study of the uncertainties in this method, the stability of the emitter, peaks in the density of states not predicted by the free electron model and the resolution of the energy analyzer may limit its interest. It was also demonstrated that performing some measurements at different temperatures\cite{young1962progress} might be useful but it requires that the temperature dependence of the physical parameters does not complicate the interpretation. In ref.\onlinecite{van1966validity}, the different experimental and theoretical limits of field emission were rather well presented and several methods were proposed, for example, to determine the emission area with an impressive 2\% uncertainty. Their conclusion was "that the Fowler-Nordheim model describes the experimental results satisfactory". In the book by Modinos\cite{modinos1984secondary}, chapters 2.3 and 2.4 give a very good summary of different experimental attempts where it appears that the best estimate of the electric field and thus of $\beta$ has been done with an uncertainty as low as 3\%. All these works are a clear improvement compared to analysis based only on the I-V curve and Eq. \ref{eqFN} but it does not allow to conclude on the closeness of these measurements with the real physical values, with maybe one exception that went almost unnoticed. In ref. \onlinecite{ehrlich1978measurement}, an absolute current density measurement was performed on a tungsten <110>. The current was measured through a probe hole and the emission area was measured independently by field ion microscopy. The current density was then estimated with an uncertainty of about 70 \%. This current density was compared to the theoretical values with and without image charge. The work function was considered to be between 4.5 and 6 eV and the value of $\beta$ was obtained from the slope of the FN plot. Despite these uncertainties, it was clearly shown that the absolute measurement of the current density lied between the triangular barrier theory and the image charge potential theoretical values. It was about 5 times lower than the image charge prediction and about 30 times higher than the triangular barrier prediction. From a metrological point of view, it can be said that there is room for improvement in field emission and that many experimentalists were too optimistic about the accuracy of their method. It might even be more correct to say that field emission experiments have a low accuracy but might show sometimes a good precision for inaccurate results.

\subsection{And then came micro and nanotechnology}

With the advent of nanosciences and their numerous poorly characterized materials and geometries, the risk was rather high that field emission experiments might lead to overwhelming spurious results. A call for standardization of field emission with an extensive list of mistakes to avoid was published\cite{zhirnov2001standardization} rather early in the field but received less attention than several experimental works about supposedly exceptional nanoemitters. Probably, this battle could not be won in the absence of a good theory with a good method to analyze data. In the past years, new analytical formulae have been proposed to better describe experimental data  \cite{forbes2008call,kyritsakis2015derivation,jensen2019reformulated,zubair2018fractional,liang2008generalized,forbes2010thin,lachmann2021statistical} requiring to measure the field emission current, the voltage derivative of the current, current fluctuations or the current noise. A fully open source tool is even available\cite{KYRITSAKIS201715} that includes corrections for small radius of curvature emitters. At even smaller scale, density functional theory calculations have been performed to estimate field emission currents at the atomic level\cite{lepetit2019three,de2019modeling}. Some theoreticians tried to compare their new calculations with experimental data but the lack of availability of good data make this task complicated. Some experimentalists\cite{popov2012development,feizi2012synthesis,forbes2021pre,ayari2021does,popov2022comparison} showed that it was possible to  plot new types of almost straight lines from their data. The term almost is chosen here to be vague enough to even include an FN plot appearing to be definitely not straight\cite{feizi2012synthesis} and this is rather problematic. What useful information can be extracted from those plots remains an open question\cite{forbes2021pre,me}. Very recently, a rather versatile webtool has been presented for data interpretation\cite{ALLAHAM2022103654}, but a simple version of the MG theory was used. In a preceding article\cite{me} we showed that even for a flat emitter, some discrepancies between the different levels of approximations in standard field emission models might have an important impact on data analysis. In the next sections, we will apply these models to the analysis of field emission current and differential conductance of a simple tungsten emitter and compare their different results.

\section{Measurements on a banal tungsten emitter}

Tungsten is the most and best-studied material in field emission and so it seems rather natural to test any new theory with it. With this material, the cleaning procedure can be pushed to its limit and at the same time, like most field emission emitters, it can get contaminated very quickly and present pronounced current instabilities. Thus this makes tungsten emitter, the perfect candidate to test the robustness of a theory towards noise in real conditions.

\subsection{Experimental set up}
\begin{figure}
\includegraphics{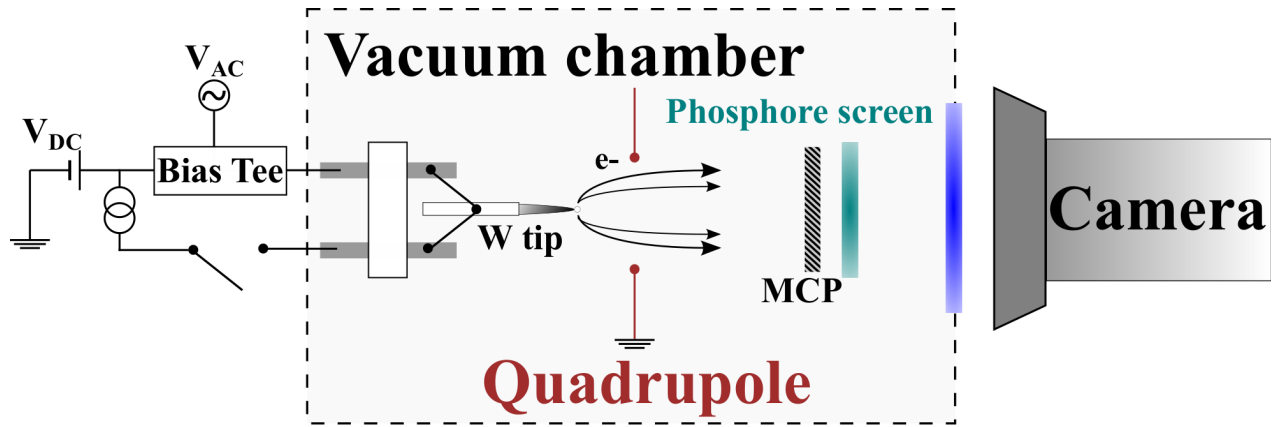}%
\caption{Experimental set-up used to perform the field emission experiments.\label{Fexp}}%
\end{figure}

A <111> tungsten tip, with an initial radius of about 15 nm, was fabricated by electrochemical etching of a 125 $\mu$m diameter tungsten wire. Field emission experiments were performed in a ultra-high vacuum chamber (see Fig.\ref{Fexp}) with a base pressure of $3\times10^{-10}$  Torr. The tip was heated, several times, for 30 seconds, with a resisting loop at a temperature of 1700 K. The tip radius is expected to increase during the experiment due to these multiple heatings. An AC voltage of 1 VRMS at 33Hz and a variable DC voltage were applied on the tip with a bias tee. A quadrupole at zero bias was in front of the tip. The current was collected with a coupled microchannel plate (MCP)/phosphor screen system connected to a homemade current amplifier and a lock-in amplifier.

The role of the quadrupole was to shield the AC signal to avoid a cross-talk between the phosphor screen and the tip by capacitive coupling. The typical cross-talk current between the tip and the quadrupole in our system is 500 pA whereas the crosstalk between the tip and the phosphor screen is below our noise floor. Applying the AC voltage on the quadrupole and recording the alternative current on the tip would require a field emission current above 100 nA to make the cross-talk negligible. We preferred to apply the AC signal on the tip and to work at a lower current to reduce the changes over time in the morphology of the emitter. Such modifications occur due to surface diffusion, ion bombardment or adsorption of residual gazes and are strongly enhanced at a higher current. If the emitter changes are too important, it is not clear if the fitting of the data could be reliably compared with any model. Another advantage of working at low DC current is to reduce the 1/f noise (where f is the frequency of the noise). Indeed, surface diffusion on the tip induces a noise current roughly proportional to the square of the DC current and with a 1/f contribution to the current spectral density. As the measurement is very sensitive to any signal offset, it is important to reduce this contribution in the current measured by the lock-in amplifier. 

The MCP/Phosphore allows visualization of any change of the emitter surface by recording the field emission pattern with a digital camera and at the same time to measure very low currents. We performed the measurements with DC field emission current in the fA to pA range. For a field emission current above 10 pA (corresponding to a current at the output of the MCP above several hundred nA), the amplification gain is not linear anymore. Before each measurement, we calibrated the offset for the DC and AC signal for zero applied DC voltage on the emitter.  The  DC offset is usually lower than 10 fA and the AC offset is lower than 10 aA.

\subsection{Current and noise measurements}\label{noisyexp}
\subsubsection{DC measurements}
\begin{figure}
\includegraphics{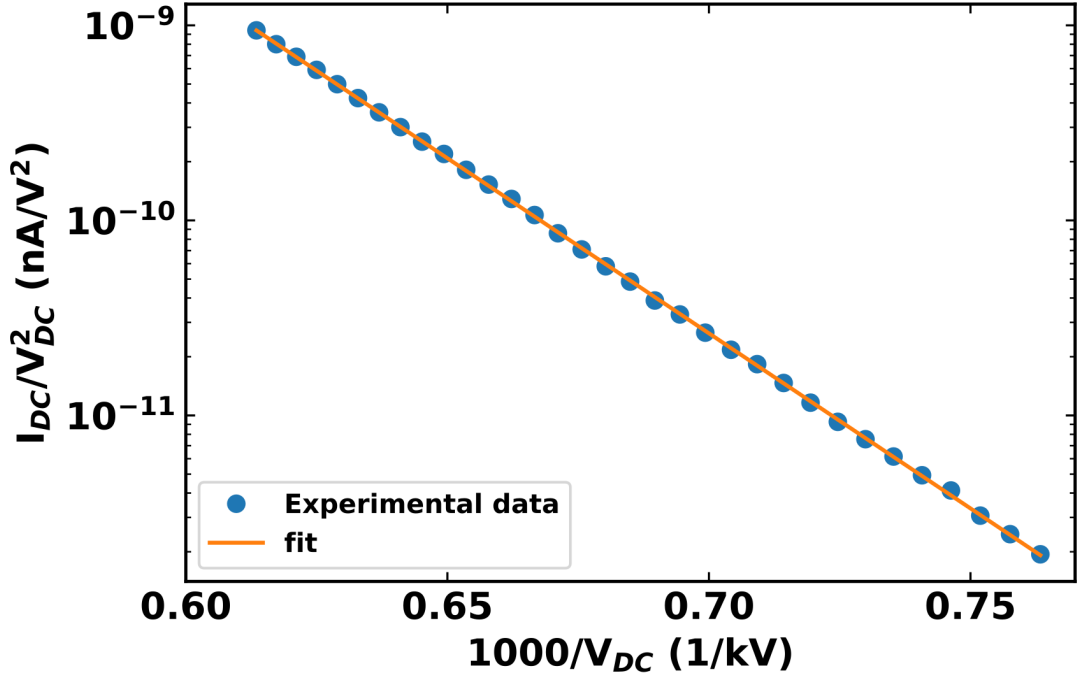}%
\caption{Typical Fowler-Nordheim plot of a <111> tungsten field emitter at a pressure of $3\times10^{-10}$  Torr. The dots are the experimental data and the solid line is a linear fit. The coefficient of determination was R$^2$ = 0.999903 and the residual = 0.0108. The fitted slope is 41339 $\pm$ 73 V. The intercept is 4.57 $\pm$ 0.05 in natural logarithm. The standard deviation of the current compared to the fit is 1.82 \% and the maximum deviation is around 6 \% (see Fig.\ref{errorI}).\label{FNexp}}%
\end{figure}
\begin{figure}
\includegraphics{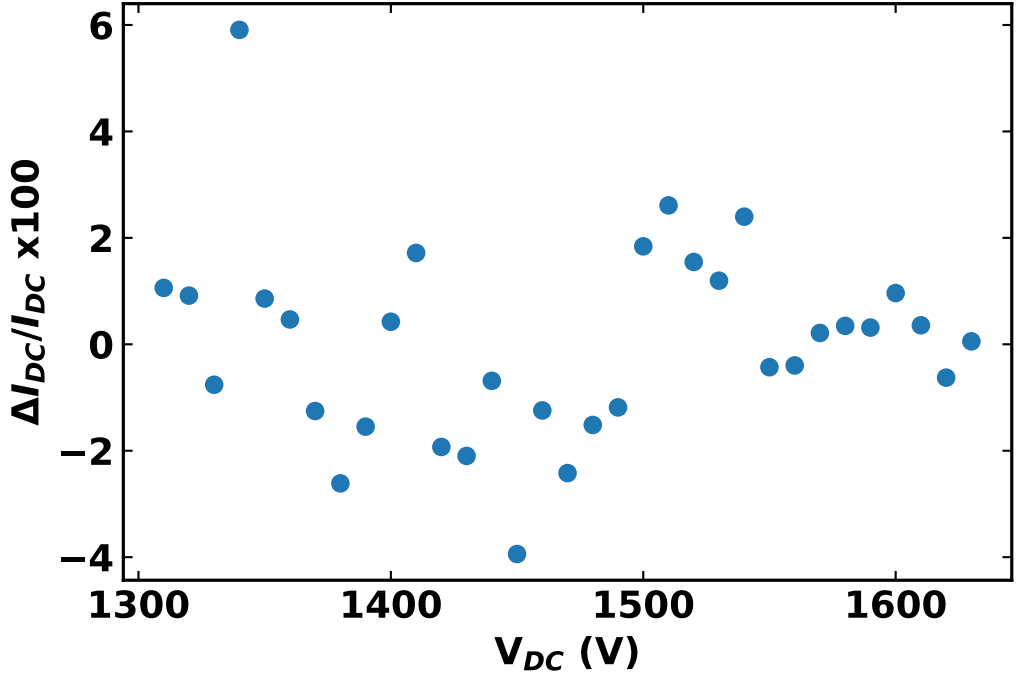}%
\caption{Relative deviation (Eq.\ref{devI}) of the experimental direct current I(V) from its fit shown in Fig \ref{FNexp}.\label{errorI}}%
\end{figure}

Several voltage sweeps between 1300 V and 1630 V were performed. We have carried out data analysis using NumPy and SciPy Python packages. The data and programs used in this article can be found in the Zenodo data repository\cite{ayarizeno}. In each case, we fitted independently the up and down sweeps of the standard Fowler-Nordheim plots of the data to a straight line (See figure \ref{FNexp}) to extract the slope $A_{FN}$ and intercept $B_{FN}$ such that : 
\begin{equation}
\label{defAB}
\log\frac{I}{V^2} = \frac{A_{FN}}{V} + B_{FN}
\end{equation}
where $\log$ is the natural logarithm. Very good coefficients of determination R$^2$ between 0.9993 and 0.9999 were obtained. The deviation $\Delta I (V)/I$ of the experimental DC current $I(V)$ to the fit $I_{fit}(V)$  (figure \ref{errorI}) was calculated for each applied voltage where:
\begin{equation}
\label{devI}
    \Delta I (V)/I =\frac{I(V)-I_{fit}(V)}{I(V)}
\end{equation} 

\begin{figure}
\includegraphics{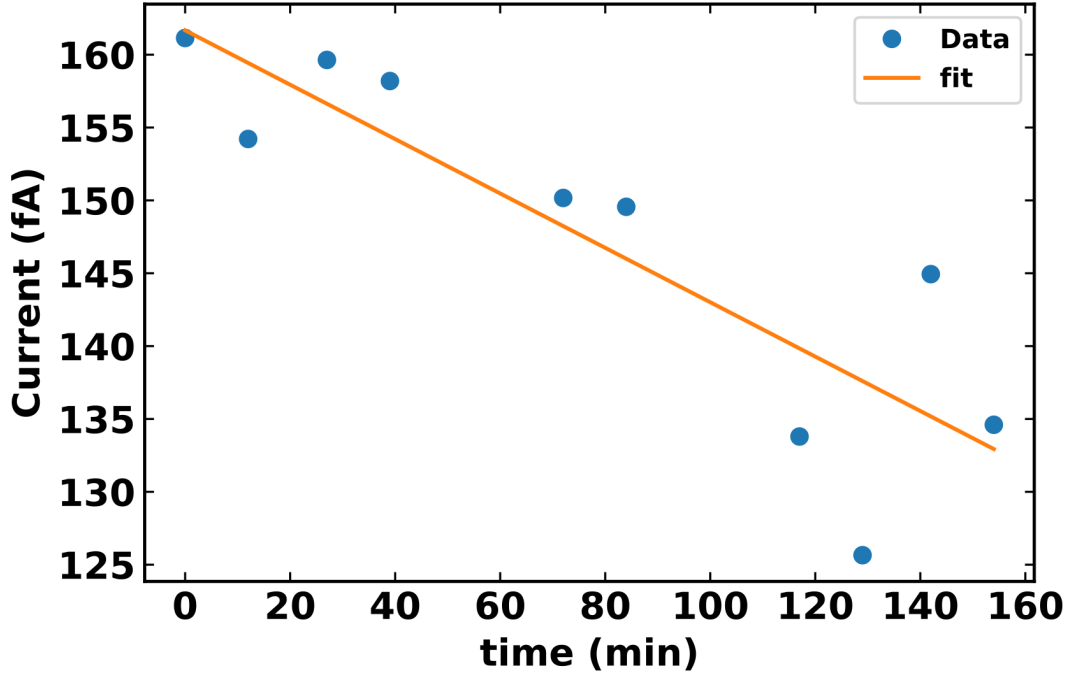}%
\caption{Field emission current as a function of time measured at an applied voltage of 1470V for each voltage sweep. The dots are the experimental data and the solid line is a linear fit.\label{SKdrift}}%
\end{figure}

Its maximum value is lower than 10 $\%$ and its standard deviation is between 1.8 and 4.4 $\%$ depending on the voltage sweeps. In this article, we plot only the data from the voltage sweep showing the lowest standard deviation and best R$^2$ but the other data sets give somewhat similar results. From Fig. \ref{errorI}, it might seem that for this run the emitter has a current Gaussian noise of a few percent and a maximum deviation of 6 $\%$. However, Fig. \ref{errorI} hides the presence of a current drift in the emitter. This can be observed for instance by comparing the current at the same voltage for different voltage sweeps as shown in Fig. \ref{SKdrift}. This drift is in the order of 8 $\%$ per hour which corresponds roughly to 1.5 $\%$ for the duration of a single voltage sweep. 
In standard experimental field emission articles, both values of current noises (Gaussian noise and drift) are generally not shown, whereas they are crucial to estimate the reliability of the fit of a Fowler-Nordheim plot. Instead, the current stability versus time for several series of fixed voltages is sometimes presented. This doesn't guarantee that the current will be the same and so the emitter will be identical when the initial voltage value is applied again. The sample might change quickly after applying a higher voltage value and the data of this evolution are sometimes simply removed. The noise values, we have obtained are rather common in good vacuum conditions, but it is not clear if this is good enough to perform a reliable estimate of the emitter parameters. In the next sections, we will try to answer to this point after analyzing our data and performing some simulations.

\subsubsection{Differential Current}
\begin{figure}
\includegraphics{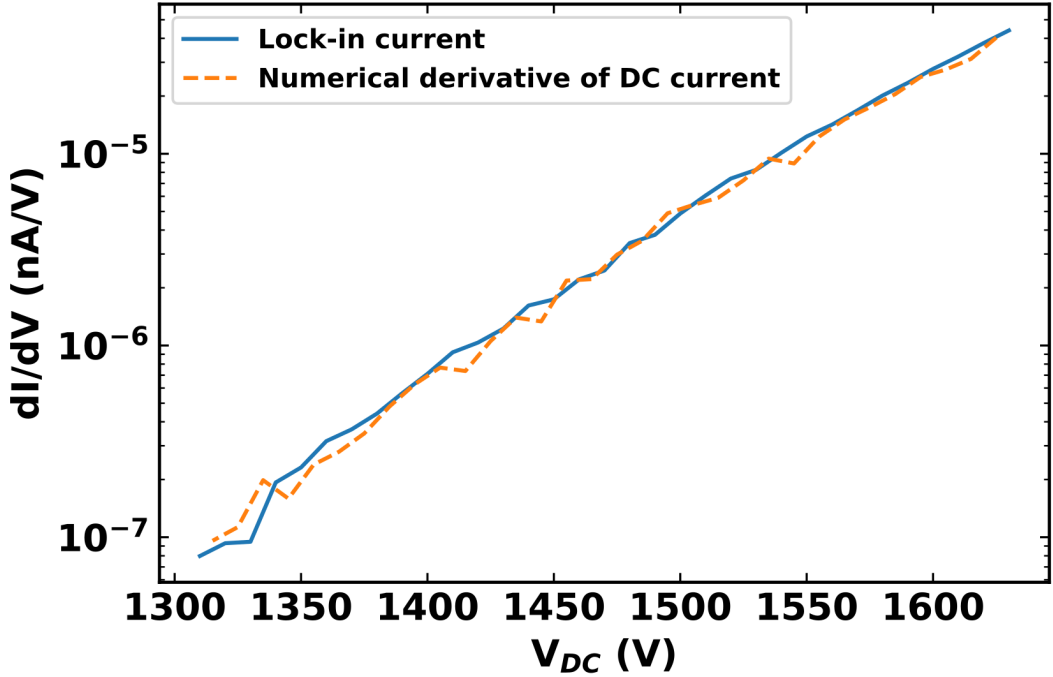}%
\caption{Lock-in current (solid line) for an applied AC voltage of 1V as a function of the applied DC voltage and numerical derivative (dashed line) for the same voltage sweep as in Fig. \ref{FNexp}.\label{Iac}}%
\end{figure}

The derivative of the current can be obtained by numerical differentiation but also with a lock-in amplifier\cite{Spindt}. A first-order Taylor expansion of the expression of the current shows that as long as the AC voltage is not too high, the current measured on a lock-in amplifier at the modulation frequency is directly proportional to the derivative of the direct current. It is usually considered that numerical differentiation methods are noisier than lock-in methods. We fitted the derivative of the current obtained by both methods and calculated their respective deviations from the fitted data in order to compare their noises. As the derivative of the current has a similar voltage dependence as the current itself, we chose to make a linear fit of the logarithm of $\partial I/\partial V\times1/V^2 $ as a function of $1/V$. The standard deviation of the lock-in signal compared to the fit is between 4 and 10 $\%$ depending on the different voltage sweeps, approximately twice as big as the DC noise. The noise of the current derivative obtained by numerical differentiation is systematically bigger than the lock-in measurement with an excess noise at best 15 $\%$ higher than the standard deviation of the lock-in data. In the worst case, the noise from the numerical derivative was 4 times bigger than the lock-in noise. The case with 15 $\%$ excess noise is presented in Fig. \ref{Iac}. Although a more systematic study would be necessary, we do not consider that most of the noise in the current derivative is coming from the 1/f noise as the coefficient of correlation between the current and its derivative do not present a clear trend and varies between 0.18 and 0.68 from one voltage sweep to another. The fact that the relative lock-in noise is higher than the relative DC noise is probably due to the low amplitude of the lock-in current. For an AC voltage of the order of one volt, the lock-in current is typically between one and two orders of magnitude lower than the DC. We will see that this level of noise in the voltage derivative of the current is an important limitation to the use of advanced field emission analysis combining the current and its derivative. 

\section{Data analysis}
In this section, several analysis methods will be tested on the data presented above. The different predictions between different runs and different methods will be highlighted.
\subsection{Traditional analysis methods}
As explained above, once a convincing fit such as the one in Fig. \ref{FNexp} is obtained, an additional hypothesis is necessary to draw some conclusions. As our emitter is a <111> tungsten, it is plausible that its work function is close to 4.5 eV but other work function values will be tested too. Our goal here is first to estimate the uncertainty arising from the analysis of a single data set and then to compare it with the experimental statistical uncertainty coming from the full data set.

\subsubsection{Analysis with a fixed work function}
\label{secAn}

\begin{figure}
\includegraphics{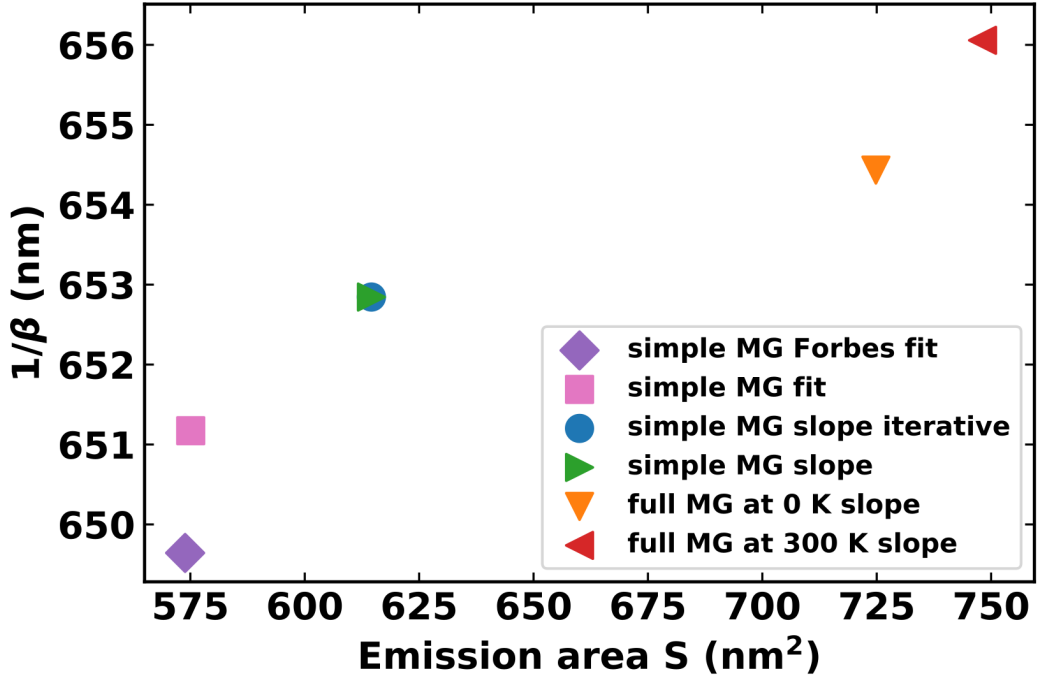}%
\caption{ Inverse of the enhancement factor $\beta$ as a function of the emission area S extracted from different methods of analysis. The method based on the triangular barrier is not represented as it predicts an area two orders of magnitude too high. The caption listed from top to bottom the different models by order of appearance in \ref{secAn}.\label{Sbeta}}%
\end{figure}

If we assume that the work function $\phi$ = 4.5 eV, we can determine the value of $\beta$ and S provided we define the field emission model we want to use. In the following, we will estimate $1/\beta$ (which gives more convenient numerical values than $\beta$) and the area from the data presented in Fig. \ref{FNexp} with seven different methods based on several field emission models. The results are gathered in Fig. \ref{Sbeta}.

The simplest model to use is the triangular barrier model with Eq. \ref{eqFN} where $\beta$ is deduced  from the negative slope of the FN plot $A_{FN}$:
\begin{equation}
\label{betaFN}
\beta = -\frac{b\phi^{3/2}}{A_{FN}}
\end{equation}
We obtained $1/\beta = 634 \pm 1$ nm. Just to have a rough idea of the size of the tip, we can use the Gomer formula\cite{gomer1955field} $\beta = 1/5r$ (see also ref. \onlinecite{de2022field} for an excellent review on the subject) where $r$ is the radius of curvature of the emitter. This radius at the apex is then equal to 126.8 $\pm 0.2 nm $. This value is sufficiently high so that the correction factor for sharp tips as proposed in ref.~\onlinecite{kyritsakis2015derivation} will not be necessary. The area of emission can be easily obtained from the intercept $B_{FN}$ of the FN plot: 
\begin{equation}
    S =  \frac{\phi}{a\beta^2}\lim_{\frac{1}{V}\to 0} \frac{I}{V^2}  = \frac{\phi}{a\beta^2}\exp(B_{FN})
\end{equation}
This area is equal to 114213 nm$^2$ although it is known that this value is much too high as will be confirmed below.

The next six methods will include the image charge correction. The current in its simplest form at 0 K is given by:
\begin{equation}
\label{IMGsimple}
    I = JS = aS\frac{(\beta V)^2}{\phi}\exp(-bv(y)\frac{\phi^{3/2}}{\beta V})
\end{equation}
where a and b are the same terms as in Eq.\ref{eqFN}, J is the current density and $v(y)$ is the barrier shape correction factor that depends on the applied electric field through the variable $y$ and can be expressed as a combination of elliptic integrals. $v(y)$ can be simplified using this approximation\cite{Forbes2006}
\begin{equation}
\label{eqForbv}
v(y) \approx 1-y^2+\frac{y^2}{3}\log(y)
\end{equation}
So for our second method, we used Eqs. \ref{IMGsimple} and \ref{eqForbv} and perform a least square fitting of the data with these equations. We obtain $1/\beta = 650$ nm and S = 573 nm$^2$. It is also possible to have an estimation of the emission half angle $\theta$ about $6^\circ$ using the Gomer formula and Eq. 14 in ref. \onlinecite{dyke1956field}:

\begin{equation}
S =  2\pi r^2 (1-\cos\theta)
\end{equation}

For the third method, we performed the same analysis but without the approximation from Eq. \ref{eqForbv} \textit{i.e.} we use the exact expression from the MG model\cite{murphy1956thermionic}. We obtained $1/\beta = 651$ nm and S = 575 nm$^2$ which are very close to the values obtained with the second model.

For the other methods, we will use the value of the slope of the FN plot instead of performing a direct non-linear fitting of the data. It is known that image charge models will not give a straight line in FN coordinates but it is anyway a common method as the deviation from a straight line is hardly visible. However, this method requires some care as the relationship between the slope of the fit and the term in the exponential in the expression of the current is not as obvious as it first seems.  The fitted slope in FN coordinates must be compared to the voltage derivative of the current and not to the term in the exponential\cite{houston1952slope}. Thus the expression of $\beta$ is now: 
\begin{equation}
\label{betaMG}
\beta = -\frac{b\phi^{3/2}}{A_{FN}}s(y)
\end{equation}
where $s(y) = v(y)-\frac{y}{2}\frac{\partial v(y)}{\partial y}$. 

For the fourth method, the value of $y$ and thus $\beta$ will be deduced self-consistently. First, a value of the voltage must be chosen to compare the slope of the FN plot to the derivative of the current at this voltage point. It seems reasonable to choose a voltage at the center of the FN plot which corresponds to the mean of the inverse of the applied voltages, but here, as it has little influence, we will prefer a voltage value $V_m$ in the center of the applied voltage range equals to 1470 V. We started the iterative process with the value of beta obtained from the preceding method. As $y$ depends on the electric field, this new $\beta$ gives $y$ and $y$ will give a new value of $\beta$ according to Eq. \ref{betaMG}. After few iterations, we obtain $1/\beta = 653$ nm. The value of $y_m$ is equal to 0.4 and $s(y_m)$ = 0.97. It is interesting to notice that $s(y)$ is close to 1 which explains why the values of $\beta$ with or without the image charge correction are very close although the value $v(y_m)$ = 0.79 has a huge influence in the values of the current density and the emission area. After some calculations, it can be shown that the emission area is given by:
\begin{equation}
    S =  \frac{\phi}{a\beta^2}\exp(b\frac{\phi^{3/2}}{\beta V_m}(v(y_{m})-s(y_{m})))\times\lim_{\frac{1}{V}\to 0} \frac{I}{V^2} 
\end{equation}
and is equal to 615 nm$^2$. It is also possible to obtain the area by dividing the measured current by the current density J from Eq. \ref{IMGsimple}. It can be calculated for each voltage point and averaged. It gives an emission area of 616 nm$^2$. So both variants to get the area are equivalent and only the first one will be used in the rest of the article.

In the fifth method, instead of using an iterative method, the value of $\beta$ that satisfies this equation is solved numerically:
\begin{equation}
\label{optimslope}
|A_{FN}| = |\frac{\partial log(I/V^2)}{\partial1/V}| = V_m\left[\frac{V_m}{J(\beta,\phi,V_m)}\frac{\partial J(\beta,\phi,V_m)}{\partial V}-2\right]
\end{equation}
with J the current density in Eq. \ref{IMGsimple}. In this case, Eq. \ref{optimslope} is strictly equivalent to Eq. \ref{betaMG} but Eq. \ref{optimslope} can also be used for another expression of J. We obtained $1/\beta = 653$ nm, an identical value compared to the iterative method. The area is given by rewriting Eq. \ref{defAB}:
\begin{equation}
\label{optimslopeS}
S = \frac{V_m^2}{J(\beta,\phi,V_m)}\exp(B_{FN}+\frac{A_{FN}}{V_m})
\end{equation}
and gives also an identical value compared to the iterative method. For the last two methods, the general MG equation will be used (Eq. 19 in ref. \onlinecite{murphy1956thermionic}) as we showed\cite{me} that this equation predicts a substantially different exponent in the pre-exponential term in the current compared to Eq. \ref{IMGsimple}. So, to determine $\beta$ for the sixth method we will solve Eq. \ref{optimslope} and to determine the area Eq. \ref{optimslopeS} will be used but with the full MG equation at T = 0 K integrated numerically. For the last method, the same calculation is performed but at 300 K. We obtain $1/\beta = 654$ nm and S = 726 nm$^2$ for the sixth method at 0 K and $1/\beta = 656$ nm and S = 748 nm$^2$ for the last method at 300 K.

As can be seen in Fig. \ref{Sbeta}, the different methods with the image charge potential, extract fairly similar values of $\beta$ differing by less than 1 $\%$ but are not very consistent regarding the emission area as they differ by more than 20 $\%$. We deduced these physical effective parameters by making the hypothesis that the work function was known. As the <111> facet has a low density, it is very reactive even in UHV and thus it can quickly adsorb residual atoms which might change its work function. A possible value of $\phi$ can be 5.0 eV. So, the same analysis can be performed with this new reasonable work function, to see how the new estimated $\beta$ and S deviate from the preceding values. This alternative analysis gives essentially the same results for the emission area: for the third method, we obtained S = 583 nm$^2$ instead of 575 nm$^2$ for 4.5 eV and for the last method we obtained S = 755 nm$^2$ instead of 748 nm$^2$ for 4.5 eV. Of course, the extracted values of $\beta$ change significantly as the work function changes by 10 $\%$. It might be deduced from the dispersion of the data in Fig. \ref{Sbeta}, that analysis with a more accurate model tends to predict bigger $1/\beta$ and S. To the contrary, a more reliable optimization protocol that consists of using a direct least square fitting of the data instead of relying on the slope and intercept of an FN plot tends to predict lower $1/\beta$ and S.

\subsubsection{Analysis with a fixed method}\label{fifthmethod}

Although we consider that the choice among the different analysis methods is a matter of habit and a compromise between the acceptable level of complexity versus the required accuracy, we will decide somewhat arbitrarily to use the fifth method based on the simple MG equation (Eq. \ref{IMGsimple}) and the slope of the FN plot. It gives predictions in the middle range of the other methods and it is close to what is probably the most used method among the experimentalists that accept to go beyond the oversimplified triangular barrier model.

Before the experiment, the tip radius was estimated to be 15 nm from scanning electron microscope (SEM) imaging, but the value we deduced from the preceding analysis was around 130 nm. This difference can partly be explained by the heating protocol used to obtain a clean tip which leads to tip blunting. However, if we had based our analysis on considering the SEM image radius as correct and used the Gomer formula to consider $\beta$ as an input parameter, a work function of 18.86 eV would have been obtained. This result is obviously completely wrong. Such a mistake is usually done the other way around. Often the radius of curvature is overestimated (and beta is thus underestimated) for large and dirty tips that tend to form protrusions having a much smaller radius of curvature than the initial tip. This leads to an incorrect value of the work function particularly low.

\begin{figure}
\includegraphics{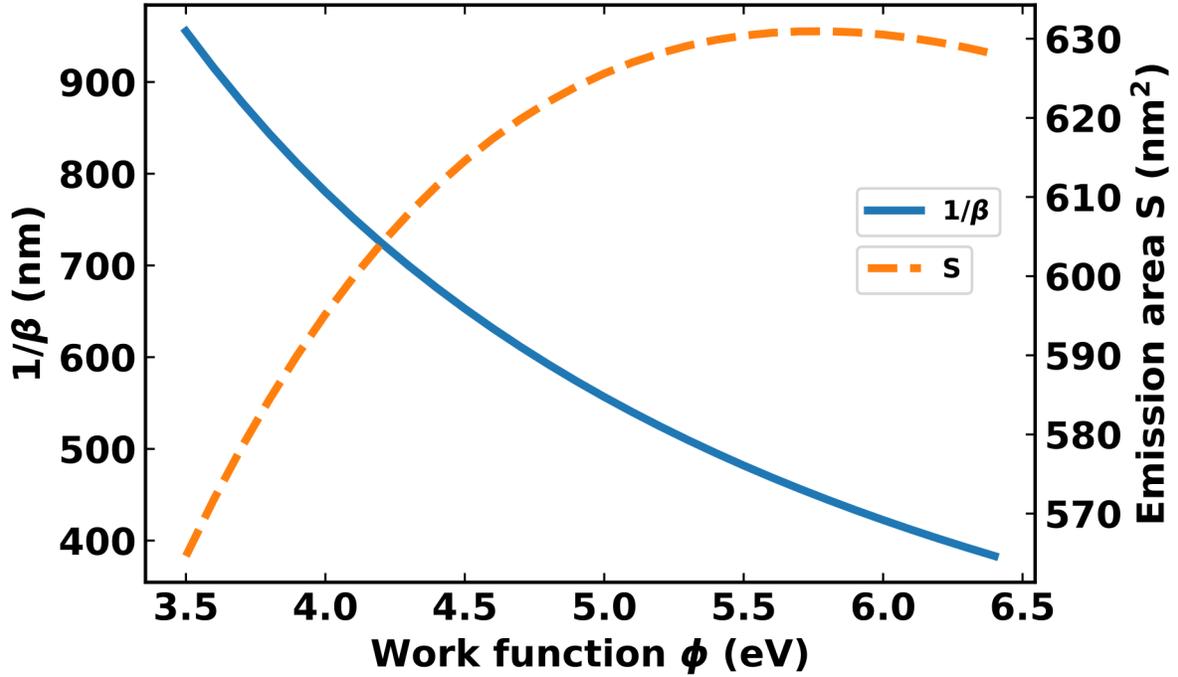}%
\caption{ Inverse of the enhancement factor $\beta$ (solid line) and emission area S (dashed line) as a function of the work function from the simple MG slope method of analysis.\label{phivar}}%
\end{figure}

Although it is not possible to extract S, $\phi$ and $\beta$ from an I-V characteristic only, it is sometimes considered\cite{charbonnier1962simple,van1966validity,forbes1999field} that it is possible to obtain a good estimate of the emission area by fitting the I-V characteristic and varying the work function over a reasonable range of physically possible values. In Fig.\ref{phivar}, such an analysis was performed on our data for a work function range between 3.5 and 6.5 eV (\textit{i.e}. 5eV $\pm 30\%$). We didn't find the same precision as in ref. \onlinecite{van1966validity}, although the variations of the estimation area are quite acceptable and lower than the dispersion we found between the different methods of analysis in \ref{secAn}. We obtained a value of the area of emission of 598 $nm^2\pm 5.5\%$. The uncertainty is slightly lower than the one mentioned in ref.~\onlinecite{forbes1999field} or ref.~\onlinecite{charbonnier1962simple} who reported respectively at best $\pm 9\%$ and $\pm 7\%$. Interestingly, even with an estimation error of the area of emission as low as $\pm 5.5\%$, the uncertainty on the 2 other parameters is still important, in particular, the error on $1/\beta$ is still $\pm 43\%$. In a way, the problem of having three physical parameters for two parameters extracted from the fit is more of a question of having two physical parameters ($\beta$ and $\phi$) for one fit parameter.

\begin{figure}
\includegraphics{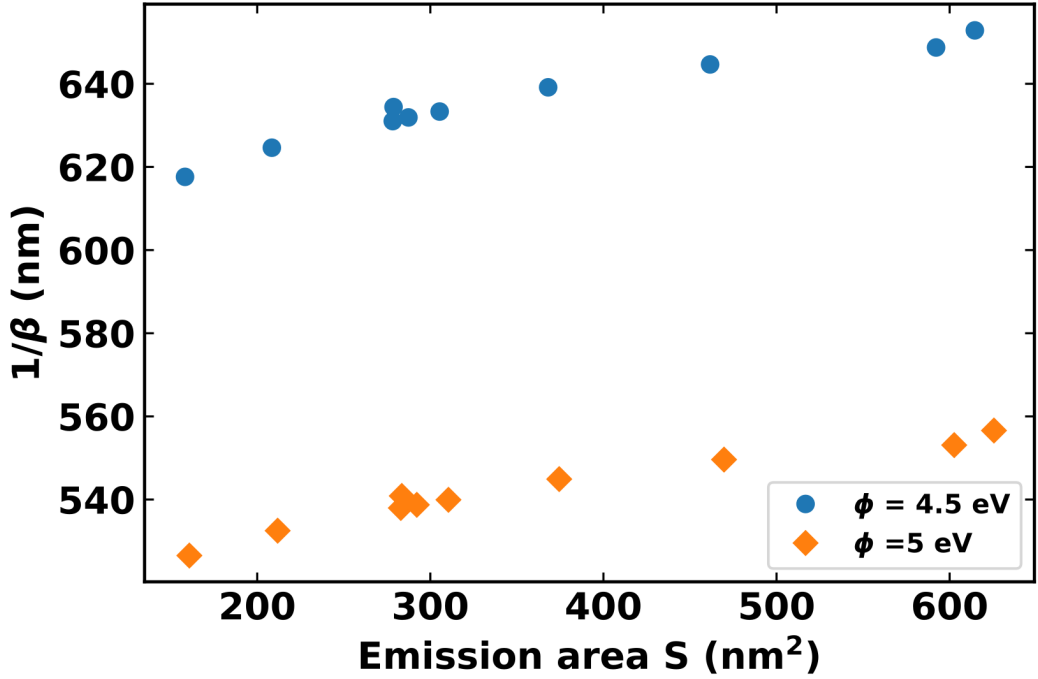}%
\caption{Inverse of the enhancement fact $\beta$ as a function of the emission area S extracted from the simple MG slope method of analysis. The dots are for a work function $\phi$ = 4.5 eV and the diamond are for 5 eV.\label{Sbetaexp}}%
\end{figure}

Now that we have an idea of the uncertainty on a single measurement voltage sweep due to the analysis, we can evaluate the uncertainty due to the evolution of the emitter. We performed the extraction of S and $\beta$ from the fifth method of analysis (MG simple slope) for two different work functions values between which the emitter work function is very likely to be (\textit{i.e.} 4.5 and 5 eV). The area between the dotted curve for $\phi$ = 4.5 eV and the diamond curve $\phi$ = 5 eV in Fig. \ref{Sbetaexp} can be considered as the zone where the emitter parameters are likely to be found. We observed a significant modification of the area of emission by a factor of 4 and the maximum change in the estimated area between 2 successive voltage sweeps can be higher than 300 nm$^2$. The experimental drift in $\beta$ is moderate and the uncertainty in the work function is the dominating issue.

It can be concluded from this section that the uncertainty in the estimation of the emission area is firstly dominated by the temporal evolution of the emitter, then by the uncertainty in the analysis method and is little influenced by the uncertainty on the work function. Regarding the uncertainty in $\beta$, it doesn't come from the uncertainty in the measurement of the current or the choice of the analysis methods but it is related to the uncertainty in the value of the work function. There is a need for a complementary measurement capable to give a reliable and independent value of $\beta$ or $\phi$. All these aspects taken into account, the uncertainty in both S and $\beta$ is of the order of $\pm 50\%$.

\subsubsection{Advanced data analysis}

\begin{figure}
\includegraphics{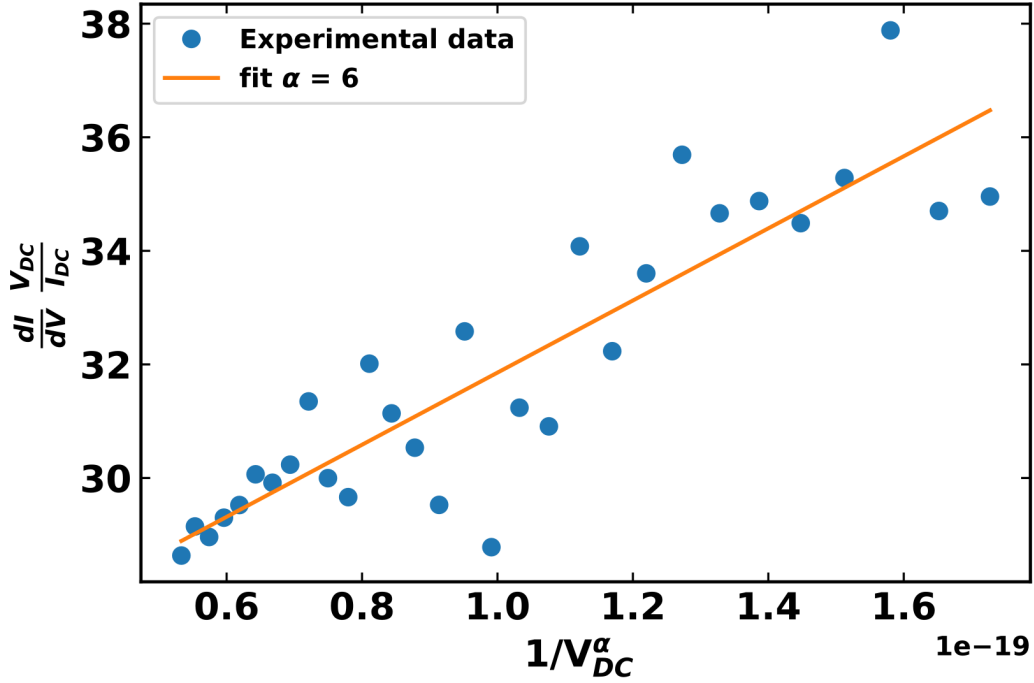}%
\caption{Variation of the product of the differential conductance by the resistance as a function of the inverse of the DC voltage at a power 6 corresponding to the best fit for Eq. \ref{eqAng2} for the same voltage sweep as in Fig. \ref{FNexp}.\label{ratio0}}%
\end{figure}

In the past few years, several new methods of data analysis have been proposed that could help to extract additional information from experimental data. One of these new methods\cite{zubair2018fractional} takes into account the roughness of the vacuum interface and proposes that field emission is described by: 
\begin{equation}\label{eqAng}
    I = C V^{2\alpha}\exp(-B/V^\alpha)
\end{equation}
where C and B are constant and $\alpha$ is a fractional space dimensionality parameter with $0 < \alpha \leq 1$ which can be reformulated into:
\begin{equation}\label{eqAng2}
   \frac{dI}{dV} \frac{V}{I} =2\alpha+\frac{\alpha}{V^\alpha}B
\end{equation}
Thus, plotting $\frac{dI}{dV}\frac{V}{I}$ as a function of $1/V^\alpha$ should give a straight line. Such a plot cannot be obtained so easily as $\alpha$ is unknown. Surprisingly, the authors\cite{zubair2018fractional} proposed a rather crude method, to say the least, that consists of plotting the left hand-side of Eq. \ref{eqAng2} as a function of $1/V$. They applied it even for a situation where they found an $\alpha$ of 0.4. Here, we will prefer to plot the left-hand side of Eq. \ref{eqAng2} as a function of $1/V^\alpha$ for different values of $\alpha$ and select the plot with the lowest least square fitting. An appealing aspect of Eq. \ref{eqAng2} is that for the best fit, the intercept should also give $2\alpha$. So we have a nice way of checking the consistency of the theory. We have seen that in our set-up the measurement of the derivative of the current is roughly twice noisier than for the DC. Nevertheless, the product of this differential conductance by the resistance is relatively well defined as shown in Fig. \ref{ratio0} with a roughly linear trend if the first low voltage noisy points are excluded. The values of $\alpha$ are not reliable as it gives the best least square fitting for values between -9 and 16.6. depending on the analyzed voltage sweep. The coefficient of determination $R^2$ is rather low, the best sweep gives $R^2$ = 0.83 (in Fig. \ref{ratio0}, $R^2$ = 0.79) and its lowest value was 0.3. Finally, the value of $\alpha$ obtained from the intercept shows no correlation with the $\alpha$ obtained from the least square fitting and varies between 9.7 and 27.7. So the data are probably too noisy to give a reliable value and the $\alpha$ we obtained is strongly out of the expected range between 0 and 1. 

Another method is based on a new analytical formula for the field emission current proposed by Forbes\cite{forbes2008call}. The formula was obtained for the case of a tunneling barrier with a classical image charge correction and can be expressed as:
\begin{equation}\label{eqForb}
    I = a_\kappa S\frac{(\beta V)^\kappa}{\phi}\exp(-b\frac{\phi^{3/2}}{\beta V})
\end{equation}
where $\kappa$ and $a_\kappa$ are independent of V and $\beta$, but vary with the work function. Now, a plot of the logarithm of $I/V^\kappa$ as a function of 1/V should give a straight line and a better fit of the experimental data. Again, as $\kappa$ is unknown, such an exact plot cannot be directly obtained. Moreover, according to more exact models, $\kappa$ is not that constant on a large voltage range\cite{ayari2021does} or because of the contribution of additional effects\cite{filippov2022field}. Despite, these significant drawbacks, the question of whether $\kappa$ might be a useful empirical parameter remains open. In order to estimate this $\kappa$, it was proposed to plot the logarithm of $I/V^\kappa$ as a function of 1/V with different $\kappa$ values to check which one gives the best fit to the data. We performed linear fits of the logarithm of $I/V^\kappa$ as a function of 1/V for different values of $\kappa$. The best least square fitting gave highly fluctuating values of $\kappa$ ranging from -4 to 13 depending on the voltage sweeps but very good coefficients of determination. For the data presented in Fig. \ref{FNexp}, the corresponding $\kappa$ is 5 with a coefficient of determination 0.99989. It means that although 99.989 $\%$ of the variance in the logarithm of $I/V^\kappa$ can be explained by the changes in 1/V according to the model, the best-extracted value of $\kappa$ is unreliable. A hint on the lack of confidence, we can have in this fit is that the value of $\kappa$ obtained from the least square fitting doesn't correspond to the maximum of the coefficient of determination, obtained here at $\kappa$ = 1.8.

$\kappa$ can also be estimated by measuring the voltage derivative of the current because Eq. \ref{eqForb} leads to the following expression:

\begin{equation}\label{theformule}
    \frac{V^2}{I}\frac{dI}{dV} = b\frac{\phi^{3/2}}{\beta}+ \kappa V
\end{equation}

\begin{figure}
\includegraphics{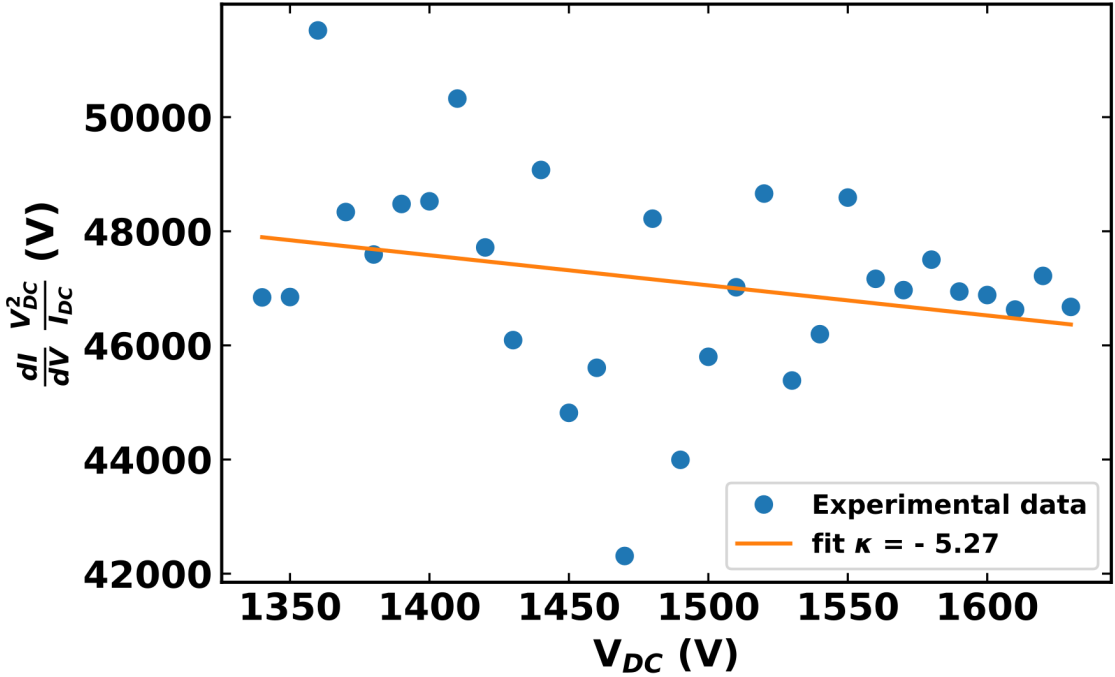}%
\caption{Variation of the expression of the left-hand side of equation \ref{theformule} as a function of the applied DC voltage. A linear fit of the curve gives $\kappa$ = -5.27 \label{ratio1}}%
\end{figure}

So a plot of this ratio as a function of V should be linear with a slope giving directly the value of $\kappa$. We found a not very convincing agreement between this expected linear behavior and our data (see Fig. \ref{ratio1}). We have performed a linear fit for different voltage sweeps and extracted from the slope a value of $\kappa$ between -11 and +15. At first sight, it seems surprising that Fig. \ref{ratio0} shows a visible trend with the voltage whereas in Fig. \ref{ratio1} the trend is drowned in the noise as they only differ by a multiplicative V term. This apparent discrepancy comes from the fact that in equation \ref{theformule}, the constant term is at least an order of magnitude bigger than the variation of the voltage-dependent term. So in Fig. \ref{ratio1}, the data represents a signal with a big intercept and a small varying term whereas in Fig. \ref{ratio0}, the data represents a signal with a small intercept and a big varying term.

Using differential conductance to extract additional information imposes stringent conditions on the acceptable level of noise. We estimated in the previous section that our uncertainty on the value of $\beta$ or S was about $\pm 50\%$ whereas in this section the fluctuations might be up to 10 times bigger than the expected value of $\kappa$. This raises the question of how much noise is acceptable to make a reliable measurement. In the last section, some simulations will be performed with a tunable level of noise to give some hints about the relation between the current noise amplitude and the uncertainty in extracted physical parameters.

\section{Numerical simulations in the presence of noise}
\begin{figure}
\includegraphics{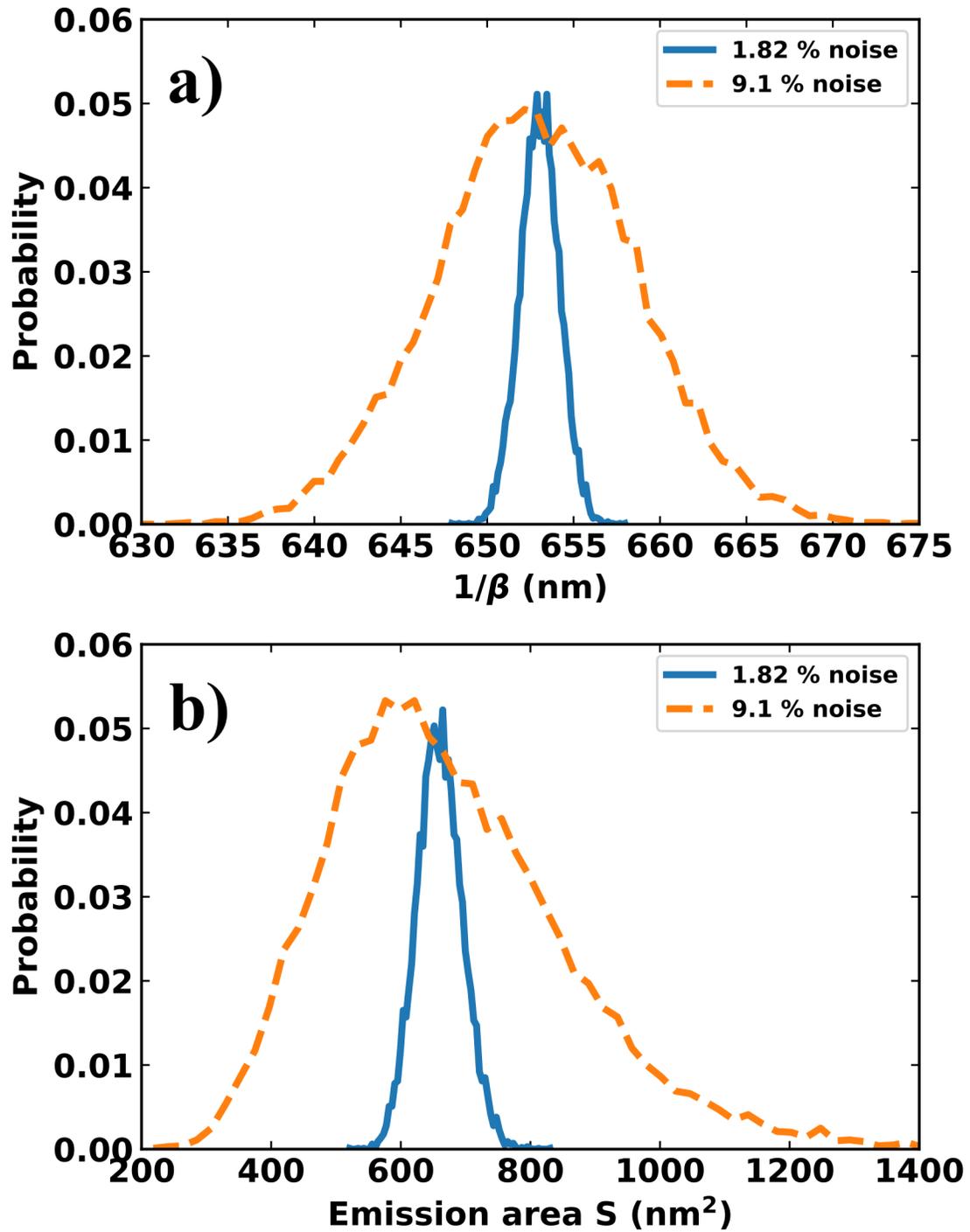}%
\caption{a) Distribution of $1/\beta$ for a noise level of 1.82 $\%$ (solid line) as in experiments and 5 times bigger (dashed line). b) Distribution for the emission for the same conditions as in a). \label{NSA}}%
\end{figure}
In \ref{noisyexp}, the current noise level was estimated with a Gaussian noise contribution and long-term drift. In the following simulations, the current drift will not be taken into account. The main reasons are i) the Gaussian noise contribution explains already the main uncertainty in the estimation of the parameters; ii) properly reproducing the drift requires a random walk approach with too many unknown parameters to model the current drift and its derivative for all voltage points. To keep it simple, we started by calculating the emission current for $\phi$ = 4.5 eV, S = 659 $nm^2$ and $1/\beta$ = 653 nm with the simple MG formula (Eq. \ref{IMGsimple}) on the same voltage range as our experiments. Then, each calculated current point is multiplied by a different randomly generated number from a normal distribution. The standard deviation of the distribution was chosen equal to 1.82 $\%$, identical to the noise value for the data shown in Fig. \ref{FNexp}. The noisy calculated current is fitted in order to extract the slope and intercept of an FN plot. $\beta$ and S are calculated by the fifth method as in \ref{fifthmethod}. This procedure is reproduced 10 000 times and a histogram of the extracted values is obtained (see Fig.\ref{NSA}). For a noise value of 1.82 $\%$, the standard deviation of the distribution of $1/\beta$ is 0.17 $\%$ and 5.3 $\%$ for the emission area. Thus, for such a level of Gaussian noise, the uncertainty induced on $\beta$ and S is not the dominant contribution and its impact is comparable to the uncertainty related to the choice of the analysis method as in Fig.\ref{Sbeta}. An increase of the noise by a factor of 5 increases the distribution width of $1/\beta$ and S by the same amount. It can be noticed however that the distribution of emission area in Fig.\ref{NSA} b) is not Gaussian anymore for such a high level of noise. The peak maximum is shifted toward a lower value but the mean is up-shifted by 20 nm$^2$. So beyond a certain noise level averaging over several experiments might lead to an overestimation of the area.

\begin{figure}
\includegraphics{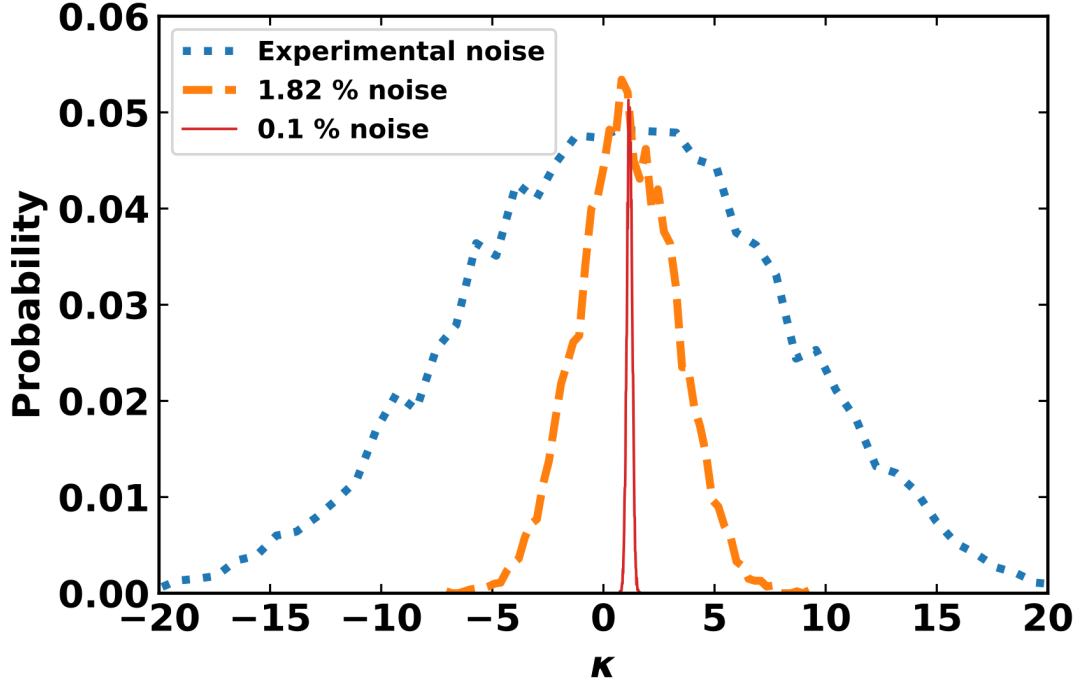}%
\caption{Distribution of $\kappa$ for a noise level similar to the data in Fig. \ref{FNexp} (dotted line), a noise level of 1.82 $\%$ (dashed line) and 0.1 $\%$ (solid line). \label{NK}}%
\end{figure}

Gaussian noise was also added to the voltage derivative of the current and the value of $\kappa$ was obtained by the same method as in Fig. \ref{ratio1}. The simulations were performed 10 000 times for three different noise levels: i) a current noise with a standard deviation of 1.82 $\%$ and noise of the derivative of the current with a standard deviation of 8.347 $\%$ as in the experimental data presented in the figures \ref{FNexp}, \ref{Iac} and \ref{ratio1}; ii) a current noise and voltage derivative of the current noise both equal to 1.82 $\%$; iii) a current noise and voltage derivative of the current noise both equal to 0.1 $\%$. In Fig. \ref{NK}, it appears clearly why in our experimental conditions the value of $\kappa$ had a huge uncertainty as the simulated $\kappa$ = 1.2 $\pm 7$ ($\approx\pm 550\%$). As the averaged value is close to the expected value, it can be argued, that performing several measurements might converge toward a reliable measurement. However, with such a level of noise, it would require about 3000 measurements to have an uncertainty below 10 $\%$. Even if a single measurement was as short as one second, the current drift and thus the changes in the emitter properties might be too important. In the case where the noise in the differential conductance is reduced down to the same value as the current noise (\textit{i.e.} 1.82 $\%$), $\kappa$ = 1.2 $\pm 2$ which is still too high. It requires to have both noises as low as 0.1 $\%$ to be in acceptable conditions where $\kappa$ = 1.18 $\pm 0.12$. However, such a low level of noise is very hard to reach experimentally. The extraction of $\kappa$ seems to require an ultra-stable emitter in an ultra-clean UHV environment. For example, It might be better to perform such measurements from a W crystal plane with lower Flicker noise like the <110> orientation\cite{swanson1967total,swanson2009review} and in extreme UHV with a vacuum level in the 10$^{-11}$ Torr range or below. After such improvements other factors might become dominant such as shot noise, vibrations, sensitivity of the detection, heating effects or temperature drifts. All these limiting factors would require to be precisely quantified, but are beyond the scope of this article. 

\section{Conclusion}

Field emission data on tungsten single emitter were analyzed using the Fowler-Nordheim model and different Murphy-Good approximation models. We showed that slight differences between traditional analysis methods using Murphy-Good approximation can give significant differences, in the order of 10$\%$, in the estimation of the emission area S. The uncertainty on beta of the order of 50 $\%$ remains dominated by the uncertainty in the value of the work function. Without an independent measure of $\beta$ or $\phi$, there is little hope to reduce this uncertainty. This partial conclusion is not especially new, but it was maybe useful to underline it with the example of a simple emitter. In our experimental UHV conditions, the drift and fluctuations of the emitter have small consequences on the value of $\beta$ but lead to an important variation of the emission area of the order of 50 $\%$. Regarding more advanced and recent analysis methods, necessitating the combination of current and differential conductance measurements, we have observed that our results were unreliable whereas our noise level is reasonable. Numerical simulations showed that the level of Gaussian noise acceptable to obtain reliable analysis is extremely low with a standard deviation of the order of 0.1 $\%$. Such an experiment might be feasible at low temperatures, with better electronics and selecting particularly stable facets with a probe hole. 

This quest for a better agreement between theory and experiment is probably the most important intellectual challenge in field emission. At the same time, it may appear to be of little practical interest because it could lead to overly complex analytical formulas or theoretical approaches. However, it clearly deserves better attention from the whole field emission community because it is also a quest toward more stable emitters; a problem of paramount importance for applications. For example, the main reason why cold cathodes despite their higher brightness are not widely used in electron microscopy is their lack of stability.


%
%

%


\section*{Conflict of interest}

The authors have no conflicts to disclose.

\section*{DATA AVAILABILITY}

The data that support the findings of this study are openly available in Zenodo at 

https://doi.org/10.5281/zenodo.7215660, Ref. \onlinecite{ayarizeno}

\bibliography{refK}

\end{document}